\documentclass[journal]{IEEEtran}

\ifCLASSINFOpdf
   \usepackage[pdftex]{graphicx}
\else
\fi
\usepackage{amssymb}
\usepackage{bibspacing}

\usepackage{tikzpagenodes}

\begin{document}
\title{Analysis of the reliability of LoRa}
\author{Qahhar Muhammad Qadir~\IEEEmembership{(Member, IEEE) \\
		\small	Salahaddin University - Erbil, Kirkuk Road, Erbil 44001, Iraq}}
\maketitle
\begin{tikzpicture}[remember picture,overlay]
\node[align=center,text=red,text width=22cm] at ([yshift=1em]current page text area.north) {$\copyright$2020 IEEE. Personal use of this material is permitted. Permission from IEEE must be obtained for all other uses, in any current or future media, including reprinting/republishing this material for advertising or promotional purposes, creating new collective works, for resale or redistribution to servers or lists, or reuse of any copyrighted component of this work in other works. Digital Object Identifier 10.1109/LCOMM.2020.3034865};
\end{tikzpicture}%

\begin{abstract}
This letter studies the performance of a single gateway LoRa system in the presence of different interference considering the imperfect orthogonality effect. It utilizes concepts of stochastic geometry to present a low-complexity approximate closed-form model for computing the success and coverage probabilities under these challenging conditions. Monte Carlo simulation results have shown that LoRa is not as theoretically described as a technology that can cover few to ten kilometers. It was found that in the presence of the combination of signal-to-noise ratio (SNR) and imperfect orthogonality between spreading factors (SF), the performance degrades dramatically beyond a couple of kilometers. However, better performance is observed when perfect orthogonality is considered and SNR is not included. Furthermore, the performance is annulus dependent and slightly improves at the border of the deployment cell annuli. Finally, the coverage probability declines exponentially as the average number of end devices grows.

\end{abstract}


%
\IEEEpeerreviewmaketitle

\section{Introduction}
%
%
%
%

\IEEEPARstart{L}{}oRa \cite{Semtech}, Semtech's solution for low power wider area network (LPWAN) \cite{Raza2017}, \cite{Qadir2018}, among other technologies, has appeared as a strong candidate for the IoT applications. Perfect orthogonality for LoRa network, i.e. frames that are set with different spreading factors (SF) are orthogonal to each other and do not cause interference, has been assumed considering that there is at least 16dB co-channel rejection between any different SF couples \cite{Georgiou2017}, \cite{Afisiadis2020}, \cite{Goursaud2015}, \cite{Georgiou2020}. This assumption however, has not been taken by others when analyzing the performance of LoRa networks \cite{Mahmood2019}, \cite{Waret2019}, \cite{Croce2020}, \cite{Croce2018}. In order to have a more realistic scenario and include the effect of all possible interference, this letter also considers imperfect orthogonality between SFs.

LoRa has attracted the attention of researchers. \cite{Goursaud2015} has reported that two LoRa devices can use different SFs so long as there is no a significant difference in their received powers. LoRa scalability and packet outage probability have been analyzed in \cite{Georgiou2017}. Although densification of LoRa gateways contributes to higher number of Co-SF interference in the lower SF tiers, it can enhance the scalability, maintain an adequate coverage, and is expected to increase network capacity \cite{Georgiou2020}. Frame error and symbol rates were approximated through deriving low complexity formulas \cite{Afisiadis2020}. For a single-cell LoRa network under imperfect orthogonality condition, the success and coverage probabilities have been analytically analyzed \cite{Mahmood2019}, throughout of uplink has been theoretically analyzed \cite{Waret2019}, LoRa's link level performance and network capacity have been experimentally and analytically analyzed taking the frame transmission period into account \cite{Croce2020} and experimental and simulations results have shown that packet loss due to imperfect orthogonality is an issue in LoRa network \cite{Croce2018}. 

LoRa vulnerability to a number of interference justifies the need for researches that study the success probability of LoRa packet. \cite{Georgiou2017} and \cite{Mahmood2019} have studied the outage probability of LoRa frames. Exploiting the positive correlation between the signal-to-interference ratio (SIR) value and success probability, this letter presents an SIR-based analytical model to quantify the success and coverage probabilities of LoRa frames under imperfect orthogonality in the presence of different interference using stochastic geometry concepts \cite{Haenggi2012}. Our work is different from literature as it reduces the complexity of computing these probabilities under all possible interference as explained in Section \ref{sec:LoRaBasic}. \cite{Georgiou2017} did not take the LoRa imperfect orthogonality feature into account and compared to \cite{Mahmood2019}, our proposal serves as a low-complexity approximate closed-form model for computing the frame success probability in LoRa network.

The remainder of this letter is organized as follows. Basic understanding of LoRa interference and its robustness techniques are explained in Section \ref{sec:LoRaBasic}. The studied problem is formulated in Section \ref{sec:probForm}. Section \ref{sec:model} presents a model for the success probability of LoRa frame under different interference scenarios. The proposed model is evaluated using simulation in Section \ref{sec:evaluation}. Finally, Section \ref{sec:conclusion} concludes this work.

\section{LoRa: interference vulnerability and robustness}
\label{sec:LoRaBasic}

In D2D communications such as in LoRa, transmitted data is mainly not as critical as the user data since the next message can update previous lost messages. In smart metering for example, the central station can wait for the next periodic message in case of the loss of previous messages. It is however not always acceptable. Critical IoT applications such as IoT-based healthcare messages are sensitive and thus require redundant delivery.

Efficient mechanisms must be used to enable data communications possible over the wireless technologies that operate in the industrial, scientific and medical (ISM) band including LoRa. Packets are lost in such environment due to collision which eventually results in link delay excess caused by resending the packets which have previously been lost. In practice, wireless technologies use a number of mechanisms to mitigate collision including carrier sense multiple access with collision avoidance (CSMA/CA) or ALOHA protocol, compliance with local regulation policies such as listen before talk (LBT) and duty cycles implemented by LoRa. They also implement automatic repeat request (ARQ) to resend the lost packets. Additionally, there are more reasons of packets loss in LoRa technology. Packets are sent on channels that are randomly selected for each transmission. Although, the implementation of frequency hopping is considered a way to avoid interference, it can also lead to packet loss due to the fact that end devices (ED)s may hop to an already occupied channel \cite{Semtech2015}.

LoRa has gained popularity from chirp spread spectrum (CSS) (LoRa patent modulation scheme) and quasi-orthogonal settings of SF. These unique features however, could not perfectly protect LoRa signals from interference. In addition to frame loss due to signal-to-noise ratio (SNR) being below the SF specified threshold ($\theta_{SF}$), LoRa signal depending on how the SFs are set, is subject to other types of losses caused by interference. Frames, when are sent with same SF, are subject to an interference called ``Co-SF interference", as well as to another kind of interference called ``Inter-SF interference" even if they are sent with different SF. There is also a possibility of having both (Co-SF and Inter-SF) interference. The aggregate of Co-SF and Inter-SF interference is denoted as ``SF interference" in this letter. We refer interested readers to \cite{Georgiou2017}, \cite{Mahmood2019} for more details on different types of interference in LoRa network.

\section{Problem formulation} 
\label{sec:probForm}

We consider a LoRa uplink transmission of EDs which are uniformly distributed around a single gateway over a deployment area represented in a multi-annuli single cell of disk shape and area $A$ as described in Table \ref{tab:lOrASpec}. We characterize the interference in the wireless network using Poisson Point Process (PPP) $\Phi$. $N$ EDs, where $N$ is a Poisson random variable with mean $\bar{N}=\rho A$, are represented by random points in the region $\Xi \subseteq \mathbb{R}^2$ with intensity $\rho > 0$ following an inhomogeneous PPP. ED $i$ has an Euclidean distance $d_{i}$ km from the gateway within the cell radius $R$. Furthermore, each ED respects the regional 1\% regulatory duty cycle $\tau$ restriction, and communicates with the gateway over a flat Rayleigh fading channel.

An SF that may not be completely orthogonal to other SFs, is set to each ED depending on its distance from the gateway by the network server. EDs use ALOHA protocol and share a single channel of 125KHz bandwidth. We further assume that both transmitter and receiver have isotropic antenna of gain equal to 1 and only LoRa EDs are operating in the ISM band, thus no interference from non-LoRa devices is expected. Sharing of SFs among EDs, will impose the possibility that two or more frames are transmitted with the same SF, thus cause Co-SF interference and possible loss of frames. Interference between frames transmitted with different SFs, i.e. Inter-SF interference is also possible. Although, Inter-SF interference is not as critical as Co-SF interference for LoRa network \cite{Georgiou2017}, \cite{Mahmood2019}, it is considered as a cause of interference in this letter.

\begin{table}[!th]
	\centering
	\caption{Technical specifications of a LoRa network, 25 byte message, BW=125KHz, $R$ radius and 6 annuli \cite{Semtech2015}, \cite{Georgiou2017} }
	\label{tab:lOrASpec}
		\begin{tabular*}{\columnwidth}
			{p{1pt}p{16pt}p{25pt}p{40pt} p{27pt} p{17pt} p{50pt}}
		\hline		
		\textbf{SF}	&	\textbf{Bitrate (kb/s)}	&	\textbf{Packet airtime (ms)} & \textbf{Transmissions per hour} & \textbf{Sensitivity (dBm)} & \textbf{SNR $\theta_{SF}$ (dBm)} & \textbf{Range (Km)}\\
		\hline		\hline
		7 		&	5.468	&	36.6  	& 98 & -123 & -6& $ \ 0-R/6$\\

		8 	&	3.125	&	64  & 56	& 	-126	& 	-9	&  $R/6-2R/6$	\\

		9 	&	1.757	&	113  & 31	& -129	& -12	& $2R/6-3R/6$	 \\

		10 	&	0.967	&	204  & 	17	& -132	& 	-15	& $3R/6-4R/6$	\\	

		11 	&	0.537	&	372  & 9	& -134.5	& -17.5	& $4R/6-5R/6$	\\

		12 	&	0.293	&	682  & 5	& -137	& -20	& $5R/6-6R/6$	\\		
		\hline		
	\end{tabular*}
\end{table}

\section{System model}
\label{sec:model}

In LoRa network, there is a possibility that EDs use same or different $SF$s as of the desired signal. Let $i=m \in \{0,1,2,...,N\}$ denote EDs and $j=k \in \{7,8,...,12\}$ denote $SF$s. Let $X_{i}(t)$ be the transmitted signal of the desired frame from ED $i$ which is subject to 1) additive white Gaussian noise (AWGN) with zero mean and variance $\mathcal{N} = N_{0}+NF+10logBW$ dBm \cite{Georgiou2017}, where $N_{0}$ is noise power density in dBm/Hz and NF is receiver noise figure in dB, 2) ``max Co-SF interference" from $n$, $n>1$ concurrent signals $X_{n}(t)$ of the same $SF$, where $X_{n^{+}}(t)$ is the strongest interfering signal among them, 3) Co-SF interference and 4) SF interference.

When $X_{i}(t)$ is subject to AWGN only, SNR can be given as $\mathcal{P}_{i} |h_{i}|^2 \ell(d_{i})/\mathcal{N}$, where $\mathcal{P}_{i}$ is the transmit power of ED $i$  in mW, $|h_{i}|^2$ is fading coefficient as an exponential independent identically distributed (iid) random variable with mean one \cite{Aljuaid2010}, and $\ell(d_{i})$ is the free space path loss of ED $i$ at $d_{i}$ distance from the gateway. From the Friis transmission equation, $\ell(d_{i})=\frac{\lambda}{(4\pi d_{i})^\eta}$ for the wavelength $\lambda$ and path loss exponent $\eta$ which is greater than 2 nearly for all kind of links. $X_{i}(t)$ survives the noise if SNR is at least as the threshold $\theta_{SF}$, i.e. condition $\mathbb{P}[SNR \ge \theta_{SF} | d_{i}]$ is met. Solving for $|h_{i}|^2$, the success probability considering the SNR only $P_{s_{SNR}}$ can be quantified as follows:

\begin{equation}
\label{eq:SNR}
P_{s_{SNR}}=\mathbb{P}\left[ |h_{i}|^2 \ge \frac{\mathcal{N}\theta_{SF}}{\mathcal{P}_{i}\ell(d_{i})} | d_{i} \right] = \exp\left(-\frac{\mathcal{N}\theta_{SF}}{\mathcal{P}_{i}\ell(d_{i})} \right)
\end{equation}

As mentioned earlier, $X_{i}(t)$ is also subject to Co-SF and Inter-SF interference. We consider two possibilities for the Co-SF interference; the received signal is 4 times stronger than the dominant signal $X_{n^{+}}(t)$ of the same SF, labeled as max Co-SF interference \cite{Georgiou2017} and stronger than the sum of all signals of the same SF, labeled as Co-SF interference. Let $\Gamma$ denote the SIR threshold required for the desired received signal to be successfully recovered. Equations \ref{eq:SIRmax}, \ref{eq:SIRCoSF} and \ref{eq:SIRboth} give the lower bound of $\Gamma$ considering max Co-SF, Co-SF and Inter-SF interference respectively.

\begin{equation}
\label{eq:SIRmax}
\Gamma_{maxCo-SF} = \frac{4\mathcal{P}_{i} |h_{i}|^2 \ell(d_{i})} {\mathcal{P}_{n^+} |h_{n^+}|^2 \ell(d_{n^+})}
\end{equation}

\begin{equation}
\label{eq:SIRCoSF}
\Gamma_{Co-SF} = \frac{\mathcal{P}_{i,j} |h_{i,j}|^2 \ell(d_{i,j})} {\sum_{m=1| m \neq i,j}^{N}  \textbf{1}_{m,j} {\mathcal{P}_{m,j} |h_{m,j}|^2 \ell(d_{m,j})}}
\end{equation}

\begin{equation}
\label{eq:SIRboth}
\Gamma_{Inter-SF} = \frac{\mathcal{P}_{i,j} |h_{i,j}|^2 \ell(d_{i,j})} {\sum_{m=1|m \neq i,k=7|k \neq j}^{N,12} \textbf{1}_{m,k} {\mathcal{P}_{m,k} |h_{m,k}|^2 \ell(d_{m,k})}}
\end{equation}

\noindent
where function $\textbf{1}_{m,j}$ indicates if $m$ EDs other than $i$ transmitting with the same $SF$ $j$ as $i$'s, function $\textbf{1}_{m,k}$ indicates if $m$ EDs other than $i$ transmitting with $SF$ other than $j$ and $h_{m,j}$ and $h_{m,k}$ are iid random variables independent of $h_{i,j}$. 

Due to the multipath effect, the envelope and phase of the received signal fluctuate over time when travels over Rayleigh fading channel. This fluctuation is due to the fact that the amplitude of the received signal is modulated by the fading amplitude $\alpha$ which is a random variable with mean square $\Omega = \overline{\alpha^2}$ and PDF $P_{\alpha}(\alpha) = \frac{2\alpha}{\Omega}\exp \left( \frac{-\alpha^2}{\Omega} \right)$, for $\alpha \ge 0$ \cite{Simon2000}. The received SIR ratio $\Gamma$ is proportionally related to the square of $\alpha$ and thus to the envelope of received signal, and its exponential PDF for a given threshold $a$ can be given by \ref{eq:PDF} \cite{HongShenWang1995}

\begin{equation}
\label{eq:PDF}
P_{\Gamma}(a)= \frac{1}{\bar\Gamma} \exp\left( \frac{-a}{\bar\Gamma} \right), \ \ \ \ \ a \ge 0
\end{equation}

\noindent where $\bar\Gamma = E[\Gamma]$

For FSK, the modulation used by LoRa, with coherent detection (i.e. no phase impairments), the rate of error $R_{e}$ can be expressed as the function of $\Gamma$, i.e. $R_{e}(\Gamma) = G \left( \sqrt{\Gamma} \right)$, where function $G(x)$ is defined by Equation \ref{eq:funG} and for $x > 0$ (i.e. $\sqrt{\Gamma} > 0$ which defines the valid lower bound of $\Gamma$ that can be received by the gateway), can be simplified to Equation \ref{eq:funGsimplified} \cite{Proakis2007}

\begin{equation}
\label{eq:funG}
G(x) = \frac{1}{\sqrt{2\pi}} \int_{x}^{\infty} \exp ^ {- \frac{y^2}{2}} dy
\end{equation}

\begin{equation}
\label{eq:funGsimplified}
G(x) \le \frac{1}{2} \exp ^ {- \frac{x^2}{2}}
\end{equation}

In order to find the outage probability $\mathbb{P}_{o}$, we average $R_{e}(\Gamma)$ over $P_{\Gamma}(a)$ by solving the integral in Equation \ref{eq:Pe} which then can be simplified by Equation \ref{eq:PeSimplified}.

\begin{equation}
\label{eq:Pe}
\mathbb{P}_{o} = \int_{0}^{\infty} {R}_{e}(\Gamma) P_{\Gamma}(a) d_{\Gamma}
\end{equation}

\begin{equation}
\label{eq:PeSimplified}
\mathbb{P}_{o} = \frac{1}{2} \left( 1- \sqrt{\frac{\bar{\Gamma}}{2+\bar{\Gamma}}} \ \right)
\end{equation}

Finally, substituting Equations \ref{eq:SIRmax}, \ref{eq:SIRCoSF} and \ref{eq:SIRboth} in Equation \ref{eq:PeSimplified}, quantifies the outage probability of LoRa frames under max Co-SF, Co-SF and Inter-SF interference respectively. The product of outage probabilities under Co-SF and Inter-SF interference represents the joint outage probability of SF interference. The success probability $P_{s}$ is then can be computed as $1-\mathbb{P}_{o}$.

\section{Model evaluation}
\label{sec:evaluation}

\begin{figure}[!th]
	\includegraphics[width=\columnwidth, height=5cm]{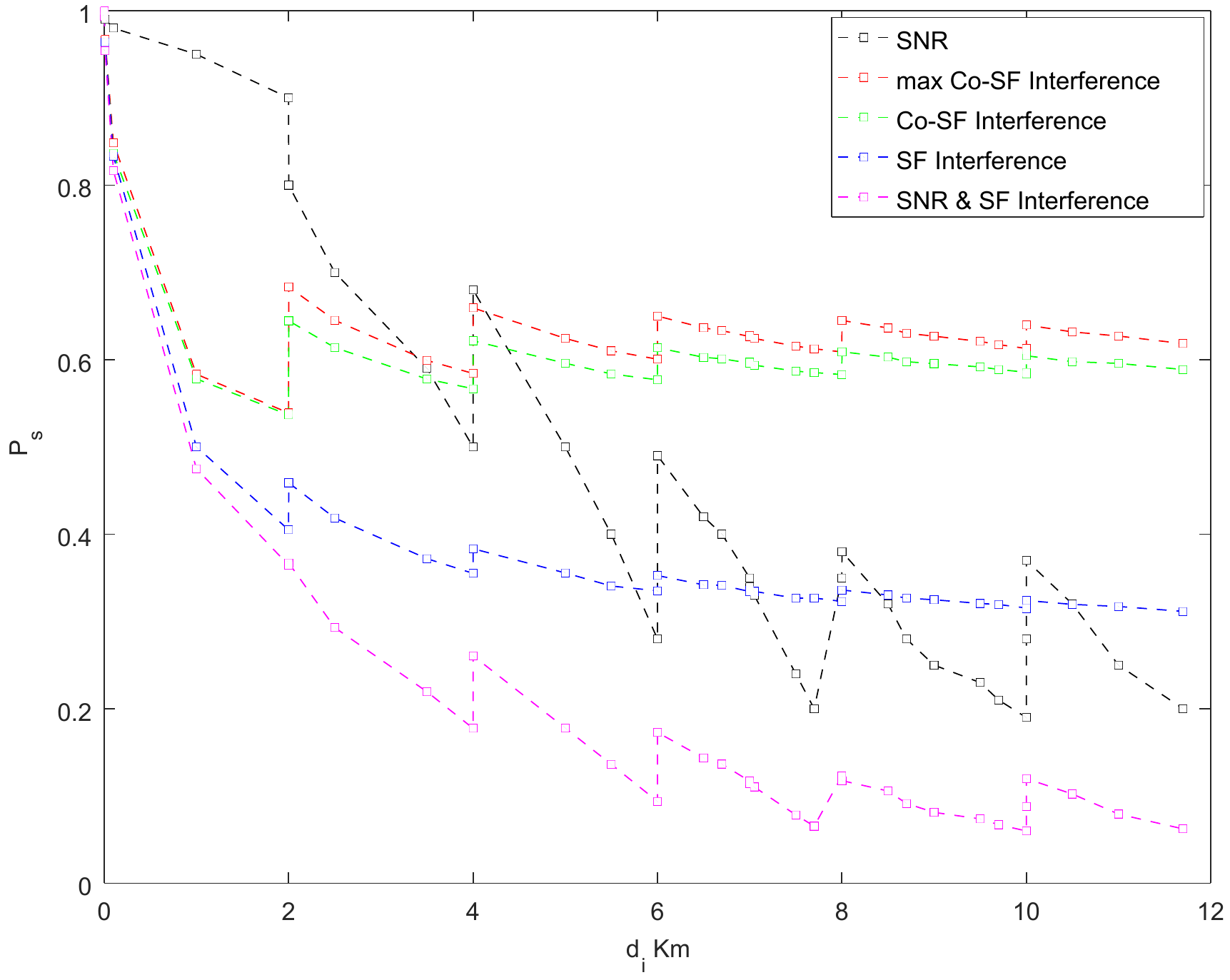}
	\caption{Success probabilities under SNR (black), max Co-SF interference (red), Co-SF interference (green), SF interference (blue) and SNR \& SF interference (magenta) versus the distance from the gateway $d_{i}$ Km, $\bar{N}$ = 1500, R = 12 Km.}
	\label{fig:sprobNoGrid}
\end{figure}

Monte Carlo simulation was used to verify the success probability in  Equation \ref{eq:SNR} and outage probability complement in Equation \ref{eq:PeSimplified}. A LoRa network of $\bar{N}=1500$ ,$R=12Km$ and six equal width ($12Km/6$) annuli as characterized by Table \ref{tab:lOrASpec}, was assumed for the simulation. Complete list of network parameters and their values used for the simulation are shown in Table \ref{tab:simParameters}. Fig. \ref{fig:sprobNoGrid} shows the averaged success probabilities under SNR, max Co-SF, Co-SF, SF, and combination of SNR \& SF interference over $10^5$ random realizations of $\Phi$ in the deployment region $\Xi$.

The overall trend of success probabilities under all interference conditions decreases as the distance from the gateway increases. Although, depending on the deployment environment, LoRa is theoretically expected to cover few to ten kilometers, its performance dramatically degrades after a couple of kilometers. However, the degradation in performance is less noticed when max Co-SF or Co-SF interference is considered, but is highly affected by noise or combination of Co-SF and Inter-SF interference. The combination of noise, Co-SF and Inter-SF interference is further degrading the performance, thus disqualifying LoRa for long ranges beyond a couple of kilometers.

The considerable difference between the performance of LoRa under Co-SF (same justification is applied to max Co-SF scenario) and combined Co-SF and Inter-SF scenario can be justified as Co-SF interference is annulus dependent due to that fact that the interference domain of EDs, using the same SF as of the desired node, is within annulus, but the interference domain of EDs in the combined Co-SF and Inter-SF scenario is within the whole deployment region $\Xi$, as in addition to interference that might be caused by EDs using the same SF as of the desired node, EDs with different SFs may also cause additional interference and thus has a combined negative impact on the performance. It is also observed that the performance is slightly improved at the border of the annuli due to the striking saw-tooth effect which is considered a LoRa unique feature. Finally, it's worth mentioning that similar trends of success probability are reported by literature \cite{Georgiou2017}, \cite{Mahmood2019}.

\begin{figure}[!th]
	\includegraphics[width=\columnwidth, height=5cm]{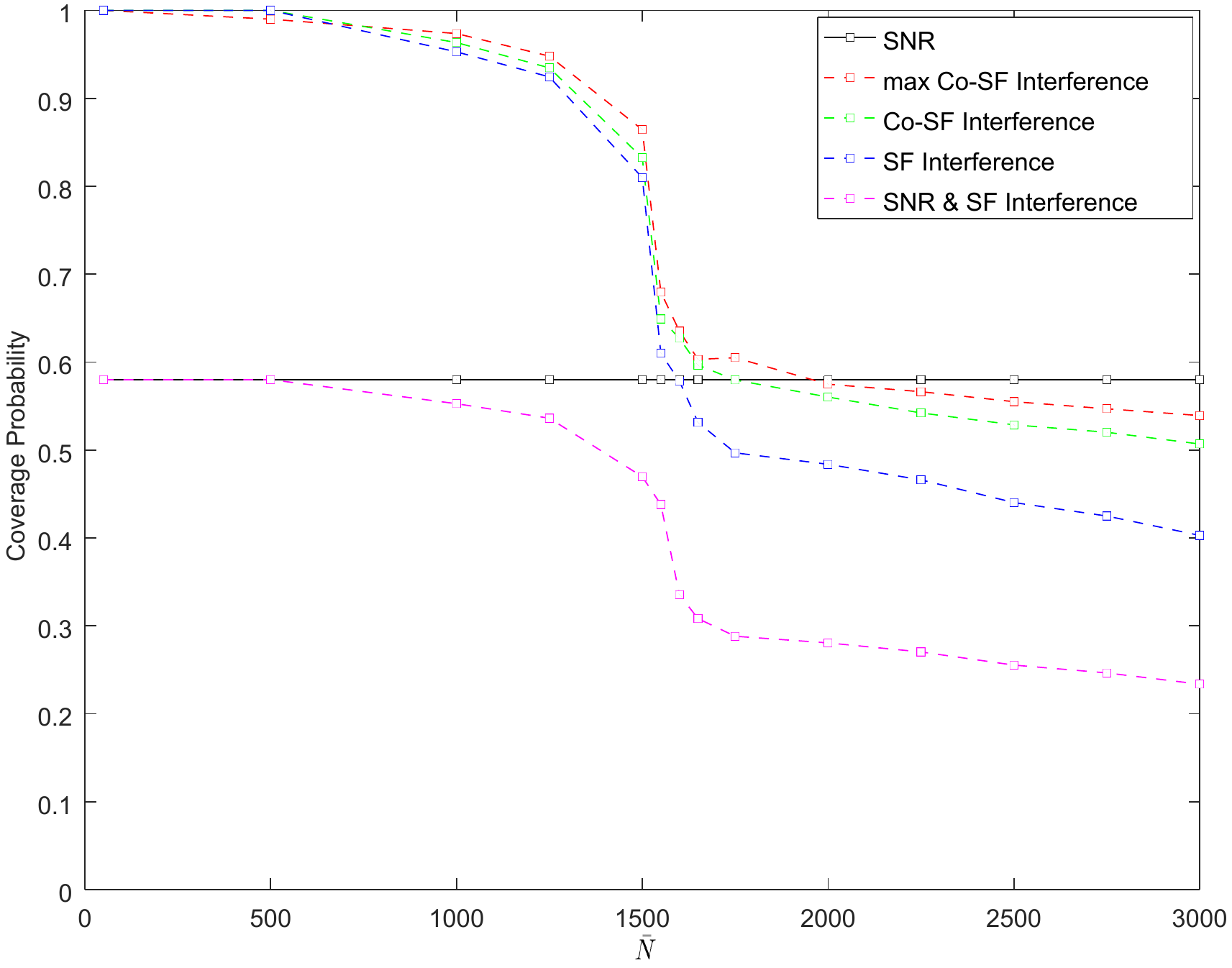}
	\caption{Coverage probabilities as functions of average number of EDs $\bar{N}\in[1,3000]$ under SNR (black), max Co-SF interference (red), Co-SF interference (green), SF interference (blue) and SNR \& SF interference (magenta).}
	\label{fig:covProbNoGrid}
\end{figure}

Finally, an exponential drop in coverage probability under considered interference conditions, is noticed as the average number of EDs grows except for the SNR, which does not depend on $\bar{N}$ (Fig. \ref{fig:covProbNoGrid}). Increasing $\bar{N}$ reduces $\Gamma$, which eventually declines the coverage probability. Sharp decay of coverage probability is observed when $\bar{N}$=1500. This point which depends on the channel condition and network settings, acts as the beginning of sever interference in the network before getting stabilized and as an indicator of different interference impact.

\begin{table}[!th]
	\centering
	\caption{Simulation parameters}
	\label{tab:simParameters}
	\begin{tabular*}{\columnwidth}
		{p{80pt}p{20pt}p{120pt}}
		\hline		
		\textbf{Parameter}	&	\textbf{Notation}	&	\textbf{Value}\\
		\hline		\hline
		Signal bandwidth & BW & 125 KHz \\
		Carrier frequency & $f_{c}$ & 868.10 MHz \\
		Noise power density & $N_{0}$ & -174 dBm/Hz \\
		Noise figure & NF & 6 dB \\
		Path loss exponent & $\eta$			&	2.7 \\
		Transmit power & $\mathcal{P}$	&	19 dBm \\
		Duty cycle		& $\tau$ &	1$\%$ \\
		Average number of EDs		& $\bar{N}$ &	1500 (Success probability) \\
		& &	$\in$ [1,3000] (Coverage probability) \\
		Cell radius		& $R$ &	12 Km \\
		Number of realizations		&  &	$10^5$ \\					
		\hline		
	\end{tabular*}
\end{table}

\section{Conclusion}
\label{sec:conclusion}
In this letter, we studied how LoRa performs in terms of frame success and coverage probabilities at the gateway in the presence of different interference considering the imperfect orthogonality between SFs. The impact of SNR, dominant Co-SF, Co-SF as well as the aggregate effect on the success and coverage probabilities, was investigated analytically and validated through Monte Carlo simulation. A model that is low-complexity approximate closed-form solution, was proposed to compute the success and coverage probabilities under each of the interference condition. 
The results have shown that the performance of LoRa is significantly affected by interference, in particular SNR, Inter-SF, and combination of all interference when the distance of EDs from the gateway exceeds a couple of kilometers. An exponential drop in coverage probability for each interference scenario was observed as the average number of EDs increases. 
An interesting area of research is to consider a multi-gateway LoRa system for further validation of the model. Furthermore, the impact of different SF combinations on $\Gamma_{Inter-SF}$ can be studied.

\ifCLASSOPTIONcaptionsoff
  \newpage
\fi



%


\end{document}